\documentclass[twocolumn,titlepage,preprintnumbers,amsmath,amssymb,aps,prl,superscriptaddress,reprint]{revtex4-1}

\usepackage{graphicx}
\usepackage{color}
\usepackage{upgreek}
\usepackage{pdfpages}
\usepackage{pgffor}
\usepackage[caption=false]{subfig}

\begin{document}

\title{Slip-Mediated Dewetting of Polymer Microdroplets}

\author{Joshua D. McGraw}
\affiliation{Soft Matter Physics Group, Experimental Physics, Saarland University, 66041 Saarbr\"ucken, Germany}
 \affiliation{D\'epartement de Physique, Ecole Normale Sup\'erieure / PSL Research University, CNRS, 24 rue Lhomond, 75005 Paris, France} 
 \author{Tak Shing Chan}
 \affiliation{Fluid Interfaces Group, Experimental Physics, Saarland University, 66041 Saarbr\"ucken, Germany}
 \affiliation{Max Planck Institute for Dynamics and Self-Organization (MPIDS), 37077 G\"ottingen, Germany} 
 \author{Simon Maurer}
 \affiliation{Soft Matter Physics Group, Experimental Physics, Saarland University, 66041 Saarbr\"ucken, Germany}
 \author{Thomas Salez}
 \author{Michael Benzaquen}
 \thanks{current address: Capital Fund Management, 23 rue de l'Universit\'e, 75007 Paris}
 \author{Elie Rapha\"{e}l}
 \affiliation{PCT Lab, UMR Gulliver 7083, ESPCI ParisTech / PSL Research University, 75005 Paris, France} 
 \author{Martin Brinkmann}
 \affiliation{Fluid Interfaces Group, Experimental Physics, Saarland University, 66041 Saarbr\"ucken, Germany}
 \affiliation{Max Planck Institute for Dynamics and Self-Organization (MPIDS), 37077 G\"ottingen, Germany}
 \author{Karin Jacobs}
\affiliation{Soft Matter Physics Group, Experimental Physics, Saarland University, 66041 Saarbr\"ucken, Germany}
\affiliation{Leibniz-Institute for New Materials, 66123 Saarbr\"ucken, Germany}

 \begin{abstract}
Classical hydrodynamic models predict that infinite work is required to move a three-phase contact line, defined here as the line where a liquid/vapor interface intersects a solid surface. Assuming a slip boundary condition, in which the liquid slides against the solid, such an unphysical prediction is avoided. In this article, we present the results of experiments in which a contact line moves and where slip is a dominating and controllable factor. Spherical cap shaped polystyrene microdroplets, with non-equilibrium contact angle, are placed on solid self-assembled monolayer coatings from which they dewet. The relaxation is monitored using \textit{in situ} atomic force microscopy. We find that slip has a strong influence on the droplet evolutions, both on the transient non-spherical shapes and contact line dynamics. The observations are in agreement with scaling analysis and boundary element numerical integration of the governing Stokes equations, including a Navier slip boundary condition. 
\end{abstract}

\maketitle

Unexpected flow phenomena emerge when the size of a liquid system is reduced below a length scale typically on the order of a few, to hundreds of nanometers~\cite{bocquetSCR09,neto05RPP, lauga07TXT}. Approaching this scale, effects associated with interfaces become increasingly important. One such effect is slip, wherein fluid slides along a solid boundary. Flow of single-component~\cite{falk10NL, whitby08NL} and complex fluids~\cite{cuenca13PRL, setu15NCOMM} in micro- and nano-channels, as well as dewetting~\cite{baeumchen14PRL, baeumchen09PRL} and interfacial instabilities~\cite{Haefner2015} of molten polymer films are systems and phenomena in which slip may have an effect. The present work demonstrates that the relaxation of micrometer-sized droplets in contact with a solid planar surface is strongly influenced by slip. The observed dewetting dynamics exhibits an unexpectedly rich phenomenology of transient droplet shapes. 

The empirical no-slip boundary condition assumes no relative motion between liquid and solid at the phase boundary. This condition was historically assumed to be valid in all practical cases~\cite{lauga07TXT}. Yet, Huh and Scriven discovered~\cite{huh71JCIS} that the no-slip boundary condition leads to infinite viscous dissipation at the tip of a liquid wedge, and thus implies that a contact line would never move -- infinite force is required to overcome infinite dissipation. In common experience we are surrounded by liquid/vapour interfaces moving along solid surfaces, from water droplets on a wind screen, to the displacement of air by liquid through a capillary or porous medium. This apparent paradox of contact line motion has attracted the attention of many researchers over at least the last four decades~\cite{huh71JCIS, dussanv74JFM, degennes85RMP, degennes03TXT, blake06JCIS, bonn09RMP, snoeijer13ARFM, sibley14JEM}.

\begin{figure*}[t!]
 \centering
\includegraphics[width=\textwidth]{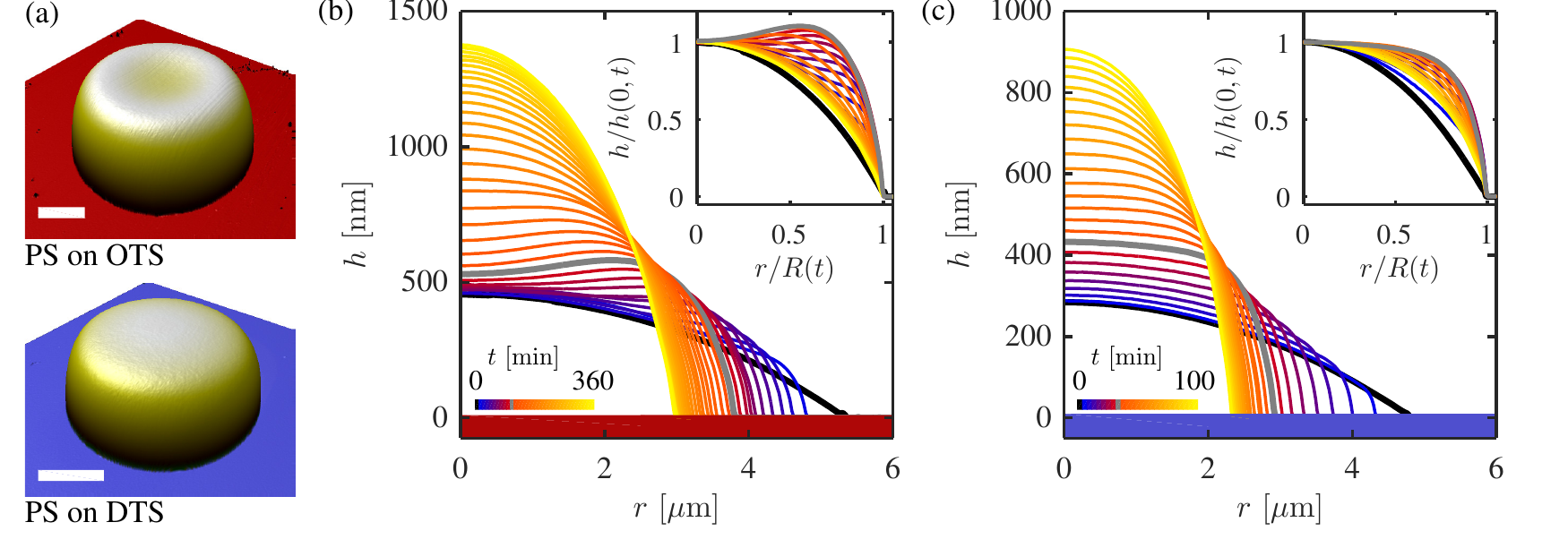}
 \caption{(a) Experimental atomic force microscopy data for 10.3 kg/mol PS microdroplets dewetting from OTS (top) and DTS (bottom) self-assembled monolayers; scale bars are $2\,\upmu$m and the height scales can be seen from the grey lines in (b) and (c). (b) Experimental height profile evolution of the PS droplet dewetting from OTS shown in (a). Time between subsequent lines is approximately 10 min. (c) Experimental height profile evolution of the PS droplet dewetting from DTS shown in (a). Time between subsequent lines is approximately 4 min. Insets show height profiles, with the radial coordinate $r$ and height profile $h(r, t)$ scaled by the contact line radius $R(t)$ and central droplet height $h(0,t)$. Inset color schemes are as in the main figures. \label{REXPT}}
\end{figure*}

Various mechanisms have been proposed to explain the existence of contact line motion. Precursor film models~\cite{degennes85RMP, heslot89JPCM, cazabat97InS, kavehpour03PRL, ghosh10ACIS, hoang11PRL, cormier12PRL, cantat14PoF} circumvent the Huh-Scriven paradox as they neglect the existence of a proper contact line. Models beyond a continuum hydrodynamic description include molecular transport mechanisms at the contact line~\cite{wayner93LAN,blake06JCIS,davitt13LAN}. Another commonly used approach to avoid the paradox is to allow for a slip boundary condition at the substrate~\cite{huh71JCIS, dussanv74JFM, thomson97SCI, vanLangerich12JFM, kirkinis13PRL, weiquing07PoF,sibley14JEM}. Importantly, all of these models employed in previous studies~\cite{marsh93PRL, davitt13LAN, delon08JFM} have in common a characteristic length scale, for example the extension of the slipping region~\cite{sibley14JEM, kirkinis13PRL} or the molecular hopping length~\cite{blake06JCIS, davitt13LAN}. This length scale is typically of nanometric size yet much smaller than the lateral extension of the interface, which could be that of a millimetric droplet or meniscus. For this wide separation of length scales, being at least five orders of magnitude, the deformation of the interface by viscous stresses is noticed only in the direct vicinity of the contact line. On a macroscopic scale, the liquid interface remains close to a quasi-static shape~\cite{eggers05PoF, chan11PoF, snoeijer13ARFM}. Millimetric sessile drops relaxing on a plane surface, for instance, are described by a sequence of spherical caps with a slowly changing apparent contact angle. In such a multi-scale system, the microscopic length has only a weak (logarithmic) effect on the dynamics~\cite{degennes85RMP, bonn09RMP}.

Recently, the no-slip hypothesis has been critically assessed. Experimental techniques~\cite{neto05RPP, lauga07TXT,bocquetSCR09} to measure slip lengths $b$, defined as the distance beyond the solid over which a linear extrapolation of the liquid velocity field reaches zero~\cite{navierOLD}, have reached nanometric resolution. Values of $b$ for small-molecule liquids on the order of tens~\cite{neto05RPP, lauga07TXT, bocquetSCR09, cho04PRL, joly06PRL,cottin08LAN, gruiyanova10MNF} up to a couple of hundred nanometers~\cite{pit00PRL, schmatko05PRL, leroyRSI09} are now reported. Polymer melts, containing chain-like molecules which can be highly coupled to one another~\cite{degennes85RMP, degenneslecture, brochard94LAN}, may show slip lengths from one up to tens of micrometers~\cite{baeumchen09PRL,Haefner2015,maxwell64PLE, reiter00LAN, leger03JPCM, fetzer05PRL, fetzer07LMR_thr, mcgraw14JCIS}. The microscopic mechanisms responsible for such large slip lengths are not clear in all cases~\cite{gutfreund13PRE}, \emph{e.g.}~for unentangled polymers dewetting from chemically similar substrates as used here~\cite{mcgraw14JCIS}. While investigations into the molecular origin of these disparate and relatively large slip lengths continue, these self-assembled monolayers (SAMs) provide an ideal set of surfaces with which to study the impact of slip on small scale interfacial flows. In contrast to systems with wide separation of length scales, a qualitatively different interfacial dynamics may be found when system sizes become comparable to the slip length.

To explore the effect of slip on small scale wetting flows, here we study polymer microdroplets dewetting from SAMs. The slip lengths involved are comparable to the typical droplet heights. Strikingly, we find that slip significantly influences the transient droplet profiles, which are non-spherical and thus not quasi-static, see~Fig.~\ref{REXPT}(a). Velocities of the receding contact lines are orders of magnitude faster than expected for no-slip systems. The good agreement between the experimental results and hydrodynamic modelling, including a Navier slip boundary condition, shows that slip is the dominating factor in the shape evolution and contact line motion of the small dewetting droplets in our experiments. Any other processes occurring at the contact line play a minor role.

The experiments were performed using spherical cap shaped polystyrene (PS) microdroplets as the initial state, with flows driven by unbalanced capillary forces. Because of the high viscosity of the non-volatile PS melt, droplets with a typical diameter of $1\,\upmu$m reach their new equilibrium conformation on the order of several minutes to hours, and can thus be monitored with high spatial resolution using atomic force microscopy (AFM). Prepared in the glassy state with a low contact angle of $\theta_0=9\pm3\,^\circ$~\cite{cormier12PRL}, the droplets are transferred onto Si wafers pre-coated with SAMs of octadecyl- or dodecyl-trichlorosilane (OTS and DTS). In the liquid state, the PS/air interfaces on these two chemically similar SAMs exhibit the same equilibrium contact angles, $\theta_\infty=62\pm3\,^\circ$. Despite the high similarity of OTS and DTS, the slip lengths of the PS melt on these SAMs are strikingly different: $b_\textrm{OTS}=160\pm 30$\,nm and $b_\textrm{DTS}=1\,500 \pm 200$\,nm are reported in Ref.~\cite{mcgraw14JCIS}. These slip lengths were extracted from the rim shape and velocity of a dewetting PS layer with an initially constant thickness~\cite{fetzer07LMR_thr,baumchen12JCM}. See the Materials and Methods section for further details on sample preparation and experimentation. 

When heated above the glass transition temperature, we observe the contact line to move inwards, while volume conservation ensures that material is collected toward the center of the droplet. Figs.~\ref{REXPT}(b) and (c) show full sequences of axisymmetric droplet profiles, $h(r,t)$ with $r$ and $t$ the radial coordinate and time, observed during the dewetting process on OTS- and DTS-covered substrates. The shapes highlighted in Fig.~\ref{REXPT}(a) were chosen such that the transient profiles deviate maximally from a spherical cap, as demonstrated in the insets (grey lines) of Figs.~\ref{REXPT}(b) and (c). 
 
The PS droplet dewetting from OTS in Fig.~\ref{REXPT}(b) transiently forms a ridge and for some time exhibits a positive curvature at the droplet center. This qualitative feature was observed previously using numerical integration of a 2D thin-film equation including a precursor film~\cite{ghosh10ACIS}. Additionally, the central droplet height does not change significantly until the width of the ridge is comparable to the time dependent contact line radius, $R(t)$ (Supporting Information, Fig. S1). In contrast to the evolution on OTS, the PS droplet dewetting from DTS does not show a pronounced ridge. Furthermore, at the earliest accessible experimental time (several minutes), the central droplet height is already increasing on DTS as seen in Fig.~S1. At late times, for both OTS and DTS, the curvature is always negative as the droplets relax to their final spherical cap shapes with the same contact angle $\theta_\infty$. In addition to the equilibrium contact angle $\theta_\infty$, surface tension $\gamma$ and viscosity $\eta$ being identical for the PS droplets on OTS and DTS in Fig.~\ref{REXPT}, the initial contact line radii $R_0=R(0)$ and initial contact angle were similar. The slip lengths, $b$, however, differ by an order of magnitude on these two substrates. The qualitative differences in the evolution are therefore expected to originate from the different values of the dimensionless slip length $\mathcal{B} = b/R_0$, as can be seen for a similar shape transition in dewetting polymer films with slip~\cite{baeumchen09PRL, fetzer05PRL}. Indeed, a gradual disappearance of the ridge at any point in the temporal evolutions can be observed by increasing $\mathcal{B}$ for PS on OTS -- decreasing $R_0$ in this case, with identical $b, \theta_0, \theta_\infty$, a ridge can no longer be detected for $R_0 \lesssim 2.7\,\upmu$m (Supporting Information, Fig. S2). 
 
To investigate theoretically whether slip is the dominating factor that controls the dewetting of PS droplets, we computed numerical solutions of the governing fluid mechanical equations. Inertial effects are neglected since the Reynolds number ${\sf Re}=\rho \dot R R/\eta\approx10^{-19}$, involving the mass density $\rho$ and contact line velocity and $\dot R=dR/dt$, is much less than unity. The flow is controlled by a balance of viscous and capillary stresses only. Solutions to the governing Stokes equation, $\boldsymbol\nabla p = \eta\boldsymbol\nabla^2\boldsymbol{u}$, and incompressibility condition $\boldsymbol{\nabla}\cdot\boldsymbol{u}=0$ are numerically computed employing a boundary element method~\cite{pozrikidis02TXT}. Here, $p$ is the scalar pressure field, and $\boldsymbol{u}$ is the velocity field. The normal stress component at the curved PS/air interface is balanced by surface tension according to Laplace's law, which reads $2\eta\partial_{\boldsymbol{n}}u_{\boldsymbol{n}}-p = 2\gamma {\cal C}$, where ${\cal C}$ is the mean curvature of the liquid/vapor interface, and where $\partial_{\boldsymbol{n}}$ denotes the directional derivative normal to this interface. Hydrostatic contributions to the pressure, elasticity of the PS melt, and corrections of the normal stress by van der Waals forces are consistently neglected in the bulk equations. To account for slippage relative to the substrate, we impose a Navier slip condition, representing a balance of stresses parallel to the substrate. The radial velocity component $u_r$ at the substrate then satisfies: 
\begin{equation}
 \kappa u_r|_{z=0} =\eta\partial_z u_r|_{z=0}\ . 
 \label{eq:Bdef1} 
\end{equation} 
Eq.~[\ref{eq:Bdef1}] allows for a definition of the slip length, $b=\eta/\kappa$, where $\kappa$ is a constant friction coefficient. As a boundary condition on the height profile, we impose a microscopic contact angle equal to the final contact angle $\theta_\infty$ at the contact line position for all times~\cite{snoeijer13ARFM,sibley14JEM}. This constant angle assumption is not strictly correct, yet, with the small variation observed in the experimental evolution, it is a reasonable approximation (examples for $\theta(t)$ are shown in Supporting Information, Fig. S3). In the following, we refer to this as the Navier-Young Model (NYM). Given $\theta_0$ and $\theta_\infty$ as control parameters, only two independent length scales can be found in the NYM for dewetting microdroplets. The droplet size represents the first length scale, \emph{e.g.}~$R_0$, while the second one is the Navier slip length $b$. Rescaling all lengths with $R_0$ and using a dimensionless time $\mathcal{T}\equiv \gamma t \eta^{-1} R_0^{-1}$, we end up with the three independent dimensionless control parameters: $\theta_0$, $\theta_\infty$, and the rescaled slip length $\mathcal{B}$ defined above. The associated dimensionless contact line radius is $\mathcal{R}(\mathcal{T}) = R(t)/R_0$, with velocity $\dot{\mathcal{R}}=d\mathcal{R}/d\mathcal{T}$.
\begin{figure}[t!]
 \centering

\begin{minipage}{\columnwidth}%
	 \includegraphics[trim=0mm 0mm 0mm 3mm, clip]{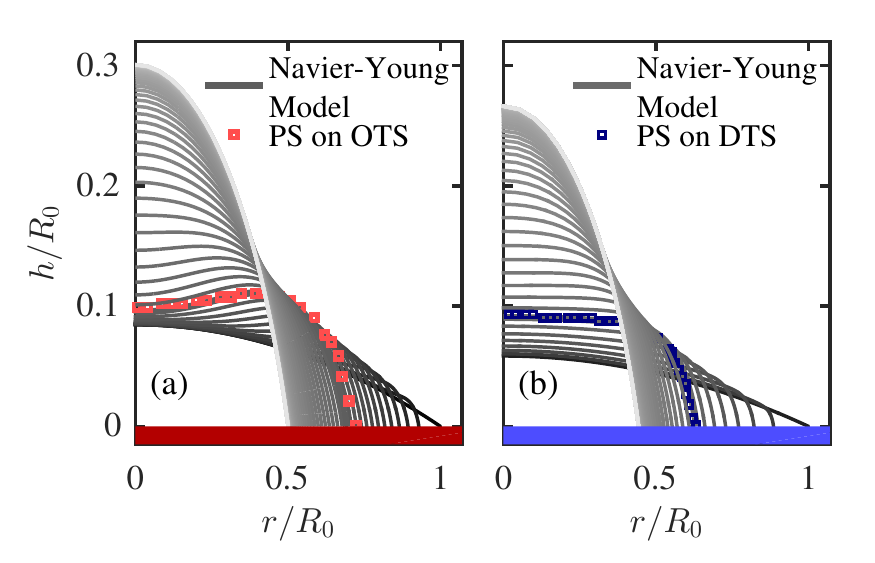}
\end{minipage}%

\begin{minipage}{\columnwidth}
	\includegraphics[width=0.95\columnwidth]{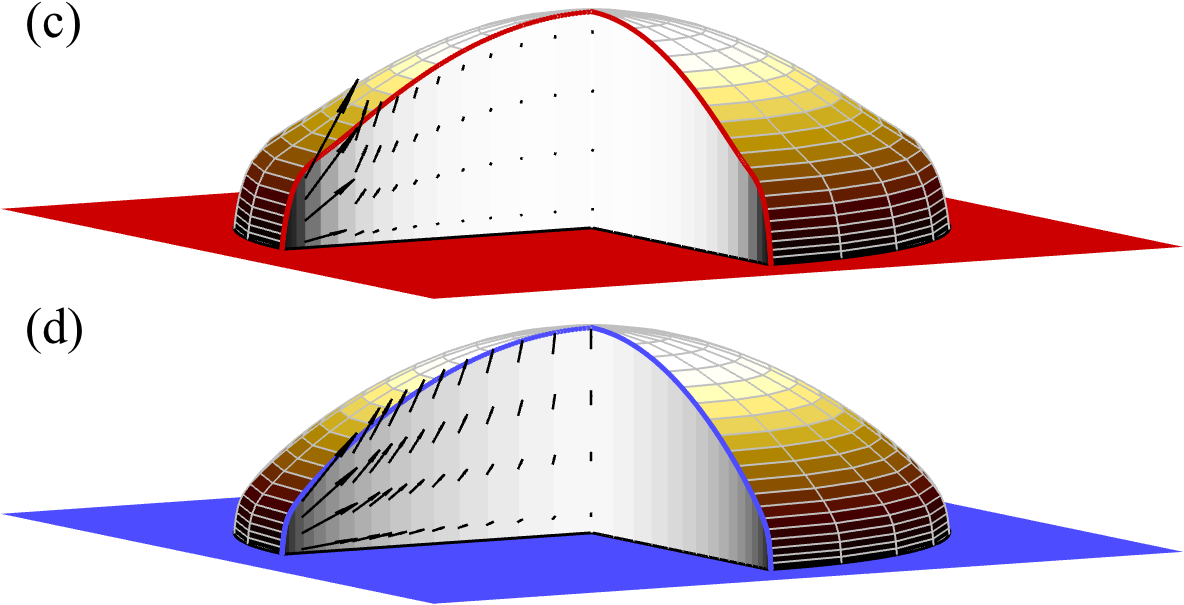}
\end{minipage}%
\caption{(a) and (b) Normalized height profiles as a function of normalized radius, for different dimensionless times, $\mathcal{T}$, for numerical droplet evolutions obtained from the Navier-Young Model: (a) $0\leq \mathcal{T} \leq 14.5$, rescaled slip length $\mathcal{B} = 0.030$, $\theta_0 = 11\,^\circ$ and $\theta_\infty = 62\,^\circ$, as for the PS droplet evolution on OTS in Fig.~\ref{REXPT}(b); (b) $0\leq \mathcal{T} \leq 5.2$, $\mathcal{B} = 0.47$, $\theta_0 = 7.0\,^\circ$ and $\theta_\infty = 62\,^\circ$, as for the PS droplet on DTS in Fig.~\ref{REXPT}(c). For both (a) and (b) experimental data from the highlighted transient profiles (grey lines of Fig.~\ref{REXPT}) are shown as squares. (c) and (d) 3D renderings of the droplet surfaces (\emph{n.b.} exaggerated vertical scales) with the flow fields when $\mathcal{R}=0.91$ for the droplets shown in (a) and (b). Grey scales indicate magnitudes of the dimensionless flow velocity for various $r/R_0$, averaged through the local interface height. Arrow lengths and grey scales are normalized by the respective contact line speeds, $|\dot{\mathcal{R}}| = 0.16$ and $1.19$ for (c) and (d). \label{BEMfig}}
\end{figure} 

In Figs.~\ref{BEMfig}(a) and (b) we present temporal profile evolutions obtained from the NYM for $\mathcal{B} = 0.030$ and $\mathcal{B} = 0.47$. The droplets have geometrical and physical parameters that were chosen to match the experimental droplets shown in Fig.~\ref{REXPT}. Remarkably, these computed droplet evolutions reproduce the curvature inversion of Fig.~\ref{REXPT}(b) and its absence in Fig.~\ref{REXPT}(c), confirming that slip plays a major role in determining the transient shape. In Fig.~\ref{BEMfig}, we also add experimental data from Fig.~\ref{REXPT} for the specific profiles that maximize the deviation from a spherical cap. Those experimental profiles show good quantitative agreement with the NYM profiles, with maximum relative deviations of $\sim5\%$. Comparison with other experiments for different droplets, as well as for different times for the droplets presented in Fig.~\ref{BEMfig}, show similar agreement. We note that the deviation between experiments and the NYM is consistent with the typical experimental uncertainties, mainly comprising overshoot due to the large slopes encountered~\cite{ioannis08TXT}, see Supporting Information, Fig.~S4. Identifying the numerically computed shapes of Fig.~\ref{BEMfig}(a) with the measured ones for the droplet on OTS of Fig.~\ref{REXPT}(b), we note that $\mathcal{B} = 0.03$ corresponds to a slip length $b_\textrm{OTS} = 160$\,nm, which is exactly the value measured in the hole-growth dewetting experiment~\cite{mcgraw14JCIS}, as given above. For the droplet of Fig.~\ref{BEMfig}(b), $\mathcal{B} = 0.47$ corresponds to a slip length $b_\textrm{DTS} = 2\, 250$\,nm for the experimental droplet of Fig.~\ref{REXPT}(c); this is similar to the measured slip length $b_\textrm{DTS} =1\,500\pm200$\,nm in Ref.~\cite{mcgraw14JCIS}.

To understand the different shape evolutions in terms of the flow structure inside the droplets, we computed the flow fields of the NYM. The numerical solutions, superimposed on the 3D renderings shown in Figs.~\ref{BEMfig}(c) and (d), reveal substantial qualitative differences in the spatial distribution of the flow inside the droplets at the early moment when $\mathcal{R}=0.91$. As illustrated in Fig.~\ref{BEMfig}(c), the velocity field for small rescaled slip length $\mathcal{B}=0.03$ is concentrated near the contact line and contains mainly shear flow. The result is an upward motion of the interface near the contact line but not in the central part of the drop. The formation of a ridge is thus observed later. In contrast to the droplet in Figs.~\ref{BEMfig}(a,c), the flow field for a larger rescaled slip length $\mathcal{B}=0.47$ in Fig.~\ref{BEMfig}(d) is more delocalized, including a significant upward flow in the central part of the droplet. We note the disparity in arrow lengths and grey scale near the droplet centers in Figs.~\ref{BEMfig}(c) and (d). The velocity field of Fig.~\ref{BEMfig}(d) corresponds more to elongational flow throughout the droplet, with the flow profile close to the contact line more resembling a plug flow (\emph{i.e.}, constant radial velocity with respect to the vertical direction). These differences between the two early flow fields are responsible for the transient ridge formation of the $\mathcal{B}=0.03$ droplet and its absence for $\mathcal{B}=0.47$. 

\begin{figure}[b!]
 \begin{center}
 \includegraphics[trim=0mm 0mm 0mm 3mm, clip]{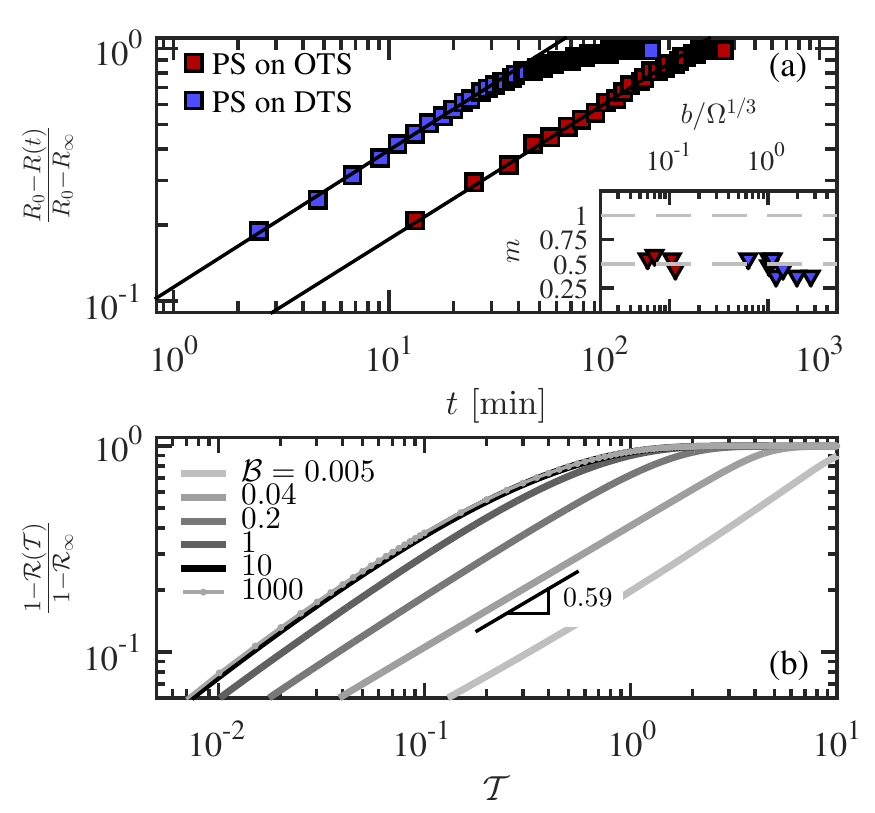}
 \end{center}
 \caption{(a) Experimental early-time dynamics of the normalized contact line radius, for the two PS droplets of Fig.~\ref{REXPT}, dewetting from OTS and DTS substrates. Solid lines represent power laws $\sim t^m$, with exponent $m = 0.54$. The inset shows such exponents measured for all studied droplets for which an early-time dynamics was accessible, as a function of $b\Omega^{-1/3}$, where $\Omega$ is the droplet volume; the dashed lines indicate $m=0.5$ and $m=1$. (b) Numerical early-time dynamics obtained from the NYM using various $\mathcal{B}=b/R_0$. Along with unit initial radius, parameters used were $\theta_0 = 9.9\,^\circ,\ \theta_\infty = 62\,^\circ$, leading to $\mathcal{R}_\infty = R_\infty/R_0 \approx 0.54$. \label{time}}
\end{figure}

Slip not only changes the structure of the flow field in the droplet, thus controlling the morphological evolution of the free interface during dewetting, but the magnitude of slip also has a significant impact on the rate of dewetting. The capillary numbers, ${\sf Ca} \equiv \dot{R}\eta/\gamma = \dot{\mathcal{R}}$, of the early flows are of order $\sim 1$ for DTS, and $\sim0.1$ for OTS, which are much larger than those typically encountered in no-slip systems~\cite{bonn09RMP, snoeijer13ARFM, davitt13LAN, marsh93PRL, delon08JFM}. To quantify the impact of slip on the dewetting dynamics, we consider the motion of the contact line. For ease of comparison, we normalize the displacement of the contact line, $R_0-R(t)$, by the total change throughout the droplet equilibration, $R_0-R_\infty$, where $R_\infty$ is the asymptotically reached contact line radius in the equilibrium state. The evolution of the normalized contact line displacement as a function of time for the two droplets of Fig.~\ref{REXPT} is shown in Fig.~\ref{time}(a), unscaled $R(t)$ are also shown in Supporting Information Fig.~S1.

In both the OTS and DTS cases of Fig.~\ref{time}, the contact line displacement during the earliest experimentally accessible time shows a power law $R_0-R(t) \propto t^m$. In addition, we show in the inset of Fig.~\ref{time}(a) the measured exponents $m$ of ten other PS droplets as a function of $b\Omega^{-1/3}$, where $\Omega$ is the droplet volume. Remarkably, although the flow structure for different cases can be significantly different, the observed value of $m$ is well represented by the average on all droplets $\langle m \rangle = 0.48 \pm 0.08$, with the error representing the standard deviation. A power law scaling of the contact line displacement in time similar to the experiments is obtained in the NYM (Fig. \ref{time}(b)) for rescaled slip length $0.03 \lesssim \mathcal{B} \lesssim 1$, corresponding to the experimental range of $\mathcal{B}$. Deviations from an ideal power law become clearly visible for $\mathcal{B} = 0.005$ and $\mathcal{B}\geq 10$. 
 
In the numerical results of the NYM for both droplets represented in Fig.~\ref{BEMfig}, we observe that the early-time frictional dissipation is concentrated near the contact line (Supporting Information, Fig. S5). At early times, we thus assume the capillary driving power to be significantly dissipated by friction in the contact line region. Since the spreading parameter~\cite{degennes03TXT} reads $S=\gamma(\cos\theta_{\infty}-1)$, the typical driving power is $\sim SR_0\dot{R}\sim -\gamma R_0\dot{R}$. On the other hand, since the frictional stress scales as $\sim\kappa\dot{R}$, the frictional dissipation power is $\sim\eta\dot{R}^2R_0\Delta/b$ (see Eq.~[\ref{eq:Bdef1}]), where $\Delta(t)$ is the typical horizontal extension of the flow region where friction is important. Supported by an argument based on a thin film approximation of the flow and by the experimentally measured profiles near the contact line (Supporting Information, Fig. S6 and text), we impose a proportionality between the vertical and horizontal extents of the slip region. Conservation of volume thus implies a scaling $R_0\Delta^2\sim R_0(R_0-R)^2$. Balancing the capillary driving power and the frictional dissipation power~\cite{nakamura13PRE} thus leads to $\dot{R}(R-R_0)\sim\gamma b/\eta$, which can be integrated into:
\begin{align}
 R_0-R(t)\sim \left(\frac{\gamma bt}{\eta}\right)^{1/2}\ ,
 \label{tauearly}
\end{align}
consistent with the values of $m$ reported above in both experiments and numerics. 
%
%

The scaling argument provided above assumes that friction at the substrate is a dominant dissipation mechanism, in addition to viscous processes~\cite{vanLangerich12JFM}. Using the NYM, we determine that friction at the substrate accounts for $\sim 60\%$ of the overall energy dissipation for $\mathcal{B}=0.03$ (OTS), and $\sim 40\%$ for $\mathcal{B}=0.47$ (DTS) during the early-time regime. While these are not necessarily dominating, it is clear that frictional dissipation is largest near the contact line where friction dominates over viscous dissipation (Supporting Information, Fig. S5). In contrast to the slip friction dominated scaling derived above, no-slip hole-growth dewetting is known to show a linear power law in time~\cite{brochard94LAN} (with logarithmic corrections). Similarly, following the arguments preceding Eq.~[\ref{tauearly}], but with the friction replaced by viscous shear dissipation and adding a regularization mechanism appropriate to the no-slip situation, a linear power law, $R_0-R(t) \sim t$, would be obtained. This shear dissipation dominated scaling prediction falls well outside our experimentally observed range of exponents in the inset of Fig.~\ref{time}(a), suggesting that dissipation in the contact line region is dominated by friction at early times.

\begin{figure}[b!]
 \begin{center}
 \includegraphics[trim=0mm 0mm 0mm 1mm, clip]{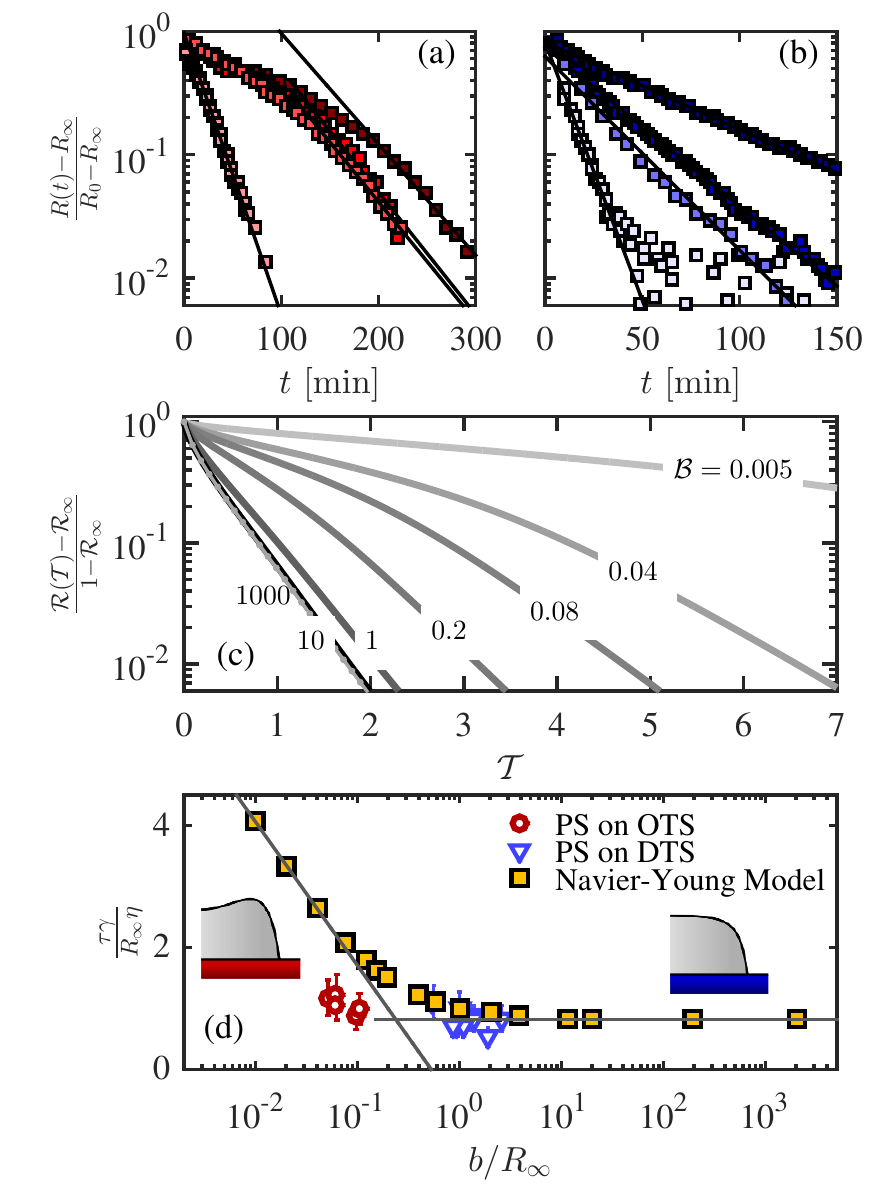}
 \end{center}
 \caption{Late-time dynamics of normalized contact line radius for: (a) PS droplets dewetting from OTS covered substrates, with $2.7 \leq R_0 \leq 5.3$\,$\upmu$m; (b) PS droplets dewetting from DTS covered substrates, with $1.8 \leq R_0 \leq 7.7$\,$\upmu$m; (c) numerical results from the NYM, with dimensionless slip lengths $\mathcal{B}$ as indicated on the curves. (d) Dimensionless relaxation time-constant as a function of dimensionless slip length, for both the experiments and the NYM. The two lines depict the asymptotic behaviours (see text): i) weak-slip logarithmic behaviour; ii) strong-slip constant behaviour. The insets recall the intermediate-time droplet shapes, for the weak-slip (red substrate) and strong-slip (blue substrate) regimes. \label{geometry}}
\end{figure}

For late-time dynamics of dewetting microdroplets, the precise value of the rescaled slip length $\mathcal{B}$ is also of great importance. In Figs.~\ref{geometry}(a) and (b), we show the displacement $R(t)-R_\infty$ of the contact line with respect to the final contact line radius $R_\infty$. As expected from a linear response, the contact line radius $R(t)$ saturates exponentially to the final equilibrium value $R_\infty$. The same features are also clearly visible in the numerical solutions shown in Fig.~\ref{geometry}(c) for various $\mathcal{B}$. Furthermore, on DTS, for the smallest droplets on OTS, and at large $\mathcal{B}$ in the NYM, the droplets reach this late-time regime at much larger $R-R_\infty$. Interestingly, an exponential relaxation to a spherical shape is also characteristic of suspended viscous droplets~\cite{rallison84ARFM, guido99JCIS}. Thus, a connection can be drawn between supported droplets with large slip lengths and free-standing droplets. In fact, by symmetry, the freely relaxing droplets are strictly equivalent to infinite-slip dewetting droplets when $\theta_\infty = 90\,^\circ$.

As for the early-time power law, the late-time exponential evolution can be understood from an energy balance. Close to equilibrium, the droplet shape is nearly a spherical cap, and the restoring capillary force is linear in $R-R_\infty$ (Supporting Information, Eq.~(S20)). The driving capillary power then scales as $\mathcal{P}_\textrm{inj} \sim-\gamma(R-R_\infty)\dot{R}$. On the other hand, the dissipation power $\mathcal{P}_\textrm{dis}$ depends on the four parameters, $\eta$, $b$, $\theta_{\infty}$, and $R_{\infty}$ and the one variable, $\dot{R}$. By dimensional analysis~\cite{barenblatt03TXT}, we can therefore write the dissipation power in the form: $\mathcal{P}_\textrm{dis}\sim\eta R_{\infty}\dot{R}^2$, with a dimensionless prefactor that is a function of $b/R_\infty$ and $\theta_\infty$ (recall that the latter is constant in this study). Equating the dissipation power with the driving one, and integrating in time leads to:\begin{align}
 R(t)-R_{\infty}&\sim\textrm{e}^{-t/\tau}\ ,\\ \tau&=\frac{\eta R
 _{\infty}}{\gamma}f\left(\frac{b}{R_{\infty}},
 \theta_\infty\right)\ ,
 \label{latepwr}
\end{align}
where $f$ absorbs the numerical prefactors missing in the scaling arguments presented above. Fig.~\ref{geometry}(d) displays the experimentally measured time constants, after a rescaling according to Eq.~[\ref{latepwr}], as a function of the dimensionless slip length $b/R_\infty$ using $b_\textrm{OTS}=160$\,nm and $b_\textrm{DTS}=2\,250$\,nm (Fig.~\ref{BEMfig}). Numerically computed time constants from the NYM allow an estimate of the scaling function $f$ in Eq.~[\ref{latepwr}], which is logarithmically  increasing for small $b/R_\infty$ but tends to a constant value for large $b/R_\infty$. Both asymptotic limits of the scaling function $f$ can be understood from the viscous dissipation in the bulk flow during the late-time relaxation. 

When $b/R_{\infty}\ll1$, the dynamics can be analyzed in the framework of the classical wedge calculation~\cite{degennes85RMP,bonn09RMP}, where the slip length replaces the microscopic cut-off length scale in the logarithmic prefactor as discussed in the introduction. In this limit, it follows for a spherical cap: $f=C_0\ln(b/R_{\infty})+C_1$, where $C_0 \approx -1.44$ for $\theta_\infty = 62\,^\circ$ (Supporting Information). In Fig.~\ref{geometry}(d), we obtain $C_0^\textrm{NYM} = -1.02$ and $C_1^\textrm{NYM} = -0.65$ by fitting the $\mathcal{B} \leq 0.08$ NYM data. While the experimental data deviate slightly (Supporting Information), the numerical results are in reasonable agreement with the analytical calculation above. 

In contrast to weak slip, when $b/R_{\infty}\gg 1$ the assumption of a localized dissipation near the contact line, inherent to the classical wedge calculation, is violated. Friction at the substrate is negligible with respect to elongational viscous stresses, and a different retraction regime is entered. In this regime, we expect $f$ to be independent of $b/R_\infty$. We find $f|_{b\rightarrow\infty} \approx0.82$ for $\theta_\infty = 62\,^\circ$ in the NYM; the experiments on DTS support this finding. The NYM value for the strong-slip relaxation time compares well with an estimate assuming viscous dissipation to occur only through elongational stresses. This simple model (Supporting Information) predicts $f|_{b\rightarrow\infty}\approx 1.05$ for $\theta_\infty = 62\,^\circ$. 

To conclude, we have studied polymer microdroplets dewetting from substrates decorated with self-assembled monolayers (SAMs). The OTS SAM provides a weaker slip boundary condition as compared to the DTS one, resulting in marked differences in the evolution of similarly sized droplets. Specifically, the weaker slip condition can give 
rise to a transient ridge. Increasing the ratio of slip length to droplet size, we observe a gradual disappearing of the ridge. These observations are explained through visualizations of the flow fields accessed through the Navier-Young Model (NYM). At early times, we find that the dewetting dynamics is in agreement with a scaling argument predicting a temporal power law evolution of the dewetted distance, with an exponent 1/2, consistent with the NYM. At late times, an exponential saturation of the contact line radius with time is observed. The time constants are in good agreement with scaling analysis and numerical solution of the NYM. This simple system of dewetting microdroplets on different surfaces gives insights on the effects of slip in free-surface micro- and nano-flows. In the context of the Huh-Scriven paradox of contact line motion, our work offers a combined experimental and theoretical justification for slip as a major control factor in the motion of contact lines.

\section*{Materials and Methods}
 To prepare the non-equilibrium droplets, PS (Polymer Standards Service GmbH) with weight-averaged molecular weight 10.3\,kg/mol, and polydispersity index 1.03, was dissolved into toluene (chromatography grade, Merck), and spin-coated onto freshly cleaved mica sheets (grade V2; Plano GmbH). After a dewetting process in toluene-saturated atmosphere at room temperature, glassy spherical cap shaped droplets on mica were produced with initial contact angles $\theta_0 = 9\pm3\,^\circ$, and contact line radii $2 \,\upmu \textrm{m} \lesssim R_0 \lesssim 7 \,\upmu \textrm{m}$ (measured using AFM; Dimension FastScan and FastScan A tips, Bruker).

Under ambient conditions, the glassy droplets were then floated from mica onto the surface of an ultraclean water bath (18 M$\Omega$\,cm, total organic carbon content $<6$\,ppb; TKA-GenPure, TKA Wasseraufbereitungssysteme GmbH), and transferred onto the SAM-coated silicon wafers ((100) crystal orientation with native oxide layer present; Si-Mat Silicon Materials). Two types of SAMs were used, being prepared from either octadecyltrichlorosilane or dodecyltrichlorosilane molecules (OTS or DTS; Sigma-Aldrich), with the self-assembly procedure and full characterization described in Ref.~\cite{lessel15LAN}. Both SAM coatings render the Si wafers hydrophobic, and lead to equilibrium PS contact angles in air of $\theta_\infty = 62\pm3\,^\circ$, as measured by AFM.

The droplets dewet when heated above the glass-transition temperature, $\approx 90\,^\circ\textrm{C}$~\cite{santangelo98MAC}. The heating stage on the AFM was set to $110\,^\circ\textrm{C}$, and AFM was used to measure \emph{in situ} height profiles of the dewetting microdroplets. This annealing temperature is low enough to ensure no loss of PS due to evaporation or degradation; volume conservation was verified. Finally, the capillary velocity of our system was measured using the stepped-film method~\cite{mcgraw12PRL} to be $\gamma/\eta = 0.07\pm0.01\,\upmu$m\,min$^{-1}$ at the experimental annealing temperature.

\begin{acknowledgments}
The authors gratefully acknowledge NSERC of Canada, the Alexander von
Humboldt Foundation, the German Science Foundation, and Total for
financial support. JDM was supported by LabEX ENS-ICFP: ANR-10-LABX-0010/ANR-10-IDEX-0001-02 PSL. 
\end{acknowledgments}



\section*{Supplementary material}

# Supporting Information for: Slip-Mediated Dewetting of Polymer Microdroplets

J.D. McGraw, T.S. Chan, S. Maurer, T. Salez, M. Benzaquen, É. Raphaël, M. Brinkmann, K. Jacobs

(Dated: March 9, 2016)

## UNSCALED CONTACT LINE RADIUS AND CENTRAL HEIGHT EVOLUTIONS

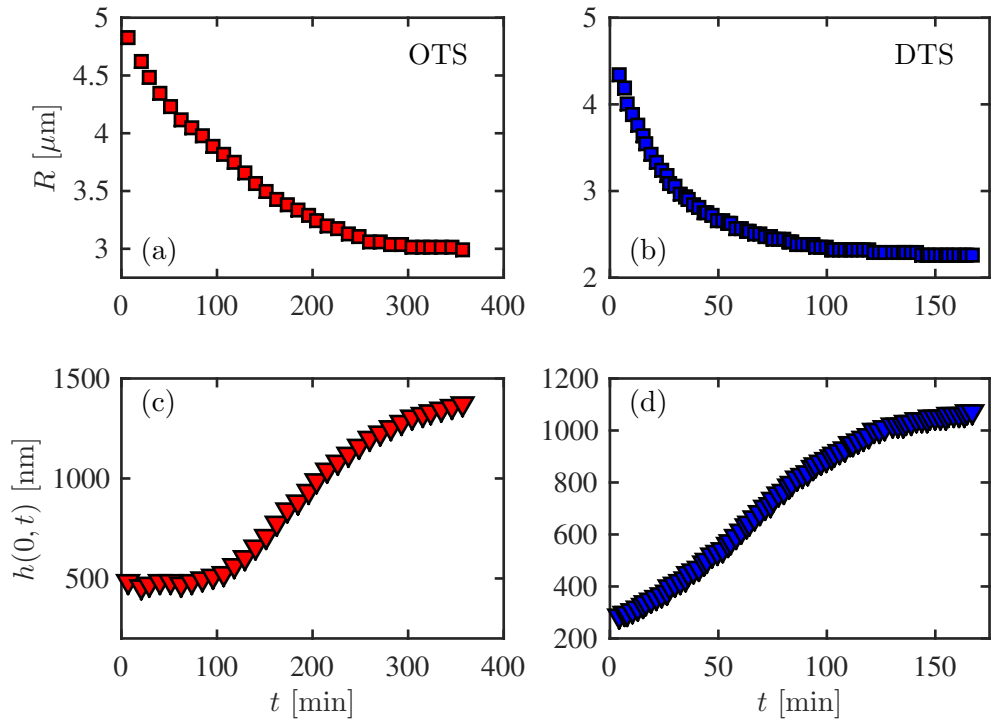

FIG. S1. Experimental contact line radius for the 10.3 kg/mol PS droplets dewetting from OTS (a) and DTS (b), as shown in Fig. 1 of the main text. (c),(d) Central height evolutions for the same droplets. Note the approximately constant central height for the droplet on (c) OTS (smaller slip, see main text) at early times, as compared to the central height for the droplet on (d) DTS (larger slip, see main text) whose height is increasing from the first experimentally accessible time.

## MAXIMAL DEVIATIONS FROM SPHERES

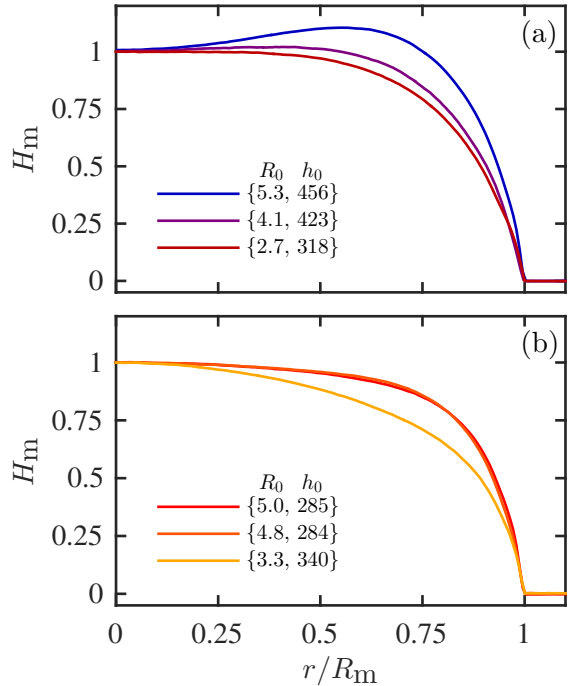

FIG. S2. (a) Experimental AFM data for PS microdroplets dewetting from OTS; here, $H_m(r/R_m) = h(r, t_m)/h(0, t_m)$, where the subscript 'm' refers to the time when the height profile achieves maximal deviation from a spherical cap (*cf.* grey lines in Fig. 1 of the main text). The legend shows the initial contact line radius, $R_0$ in μm and the initial central height, $h_0$, in nm. The ridge disappears for an initial contact line radius in the range $2.7\,\mu\text{m} < R_0 < 4.1\,\mu\text{m}$ (slip length, $b_{\text{OTS}} \approx 160\,\text{nm}$). (b) Conversely, the ridge does not appear for droplets with $R_0 \approx 5\,\mu\text{m}$ and below when dewetting from DTS ($b_{\text{DTS}} \approx 2\,250\,\text{nm}$), even while the initial contact angle, $\theta_0$, is slightly smaller in the DTS case.

## TEMPORAL EVOLUTION OF THE CONTACT ANGLE

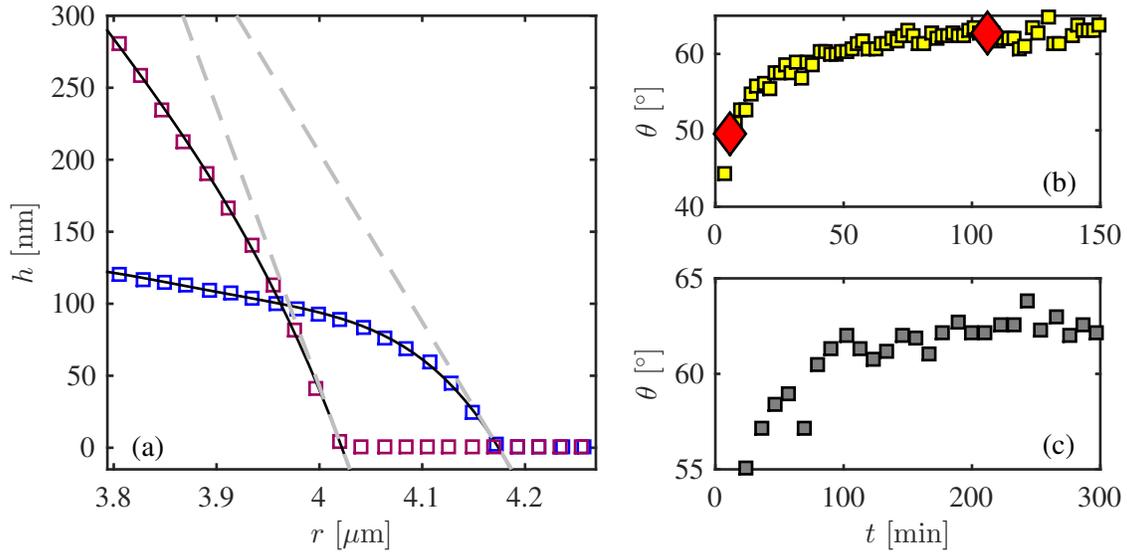

FIG. S3. (a) Experimental height profiles of a dewetting PS droplet on DTS (*cf.* Fig. 1 of the main text), from which contact angles, $\theta(t)$, can be obtained. The tangents (dashed grey lines) of a fourth order polynomial fit to the data (solid lines, not all fitted data shown) at the substrate, *i.e.* at $h = 0$, are obtained. The curve with larger contact angle has been shifted horizontally for clarity. (b) Contact angle as a function of time for the droplet represented in (a). Red lozenges indicate the curves shown in (a). We note that the initial angle (before dewetting starts) is $\theta_0 \approx 7\,^\circ$. (c) Contact angle as a function of time for the PS droplet on OTS shown in Fig. 1 of the main text.

## EXPERIMENTAL ERROR AND MODEL DEVIATIONS

We note that relative deviations between profiles predicted by the Navier-Young model (NYM) and those measured with AFM are up to 5% (Fig. 2, main text). Since the typical contact angles are rather high, $\theta \approx 60°$, overshoot can affect the absolute height measurement in the forward and reverse directions. We show an example of this overshoot in Fig. S4, where the feedback parameters that control the interaction between the AFM tip and the substrate and droplet surfaces are tuned to minimize the difference between forward and reverse traces, while allowing for a minimum tip penetration [63]. The typical deviations associated with various instrumental effects for the parameter settings of this scan (and those presented in the rest of the paper) are thus on the order of 1 %. However, independent checks on static profiles have shown that the overshoot presented here is rate dependent, and can lead to variations on the height of 5 %. Thus, the overall error on the height measurement at the 'ridge' is roughly the same as the discrepancy between experiment and the NYM. Based on the profiles alone, a discrepancy between the NYM and the experiments cannot be resolved.

While we cannot rule out other models that would change predicted profile shapes beyond the resolution of the current measurements (*i.e.* $\sim 5\%$ relative error), our numerical model with Navier slip and Young-angle boundary condition, have quantitatively corroborated the experimental height

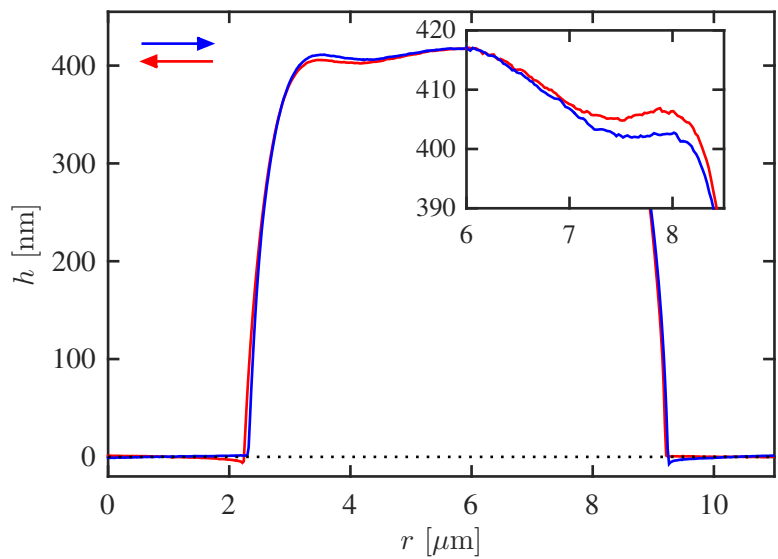

FIG. S4. Full height profiles for the forward (blue, left-to-right) and backward (red, right-to-left) traces of the AFM tip over a 10 kg/mol PS droplet dewetting from OTS at 110 °C. Contact angles as shown in Fig. S3 are only measured in directions where there is no overshoot on the substrate.

profile evolutions. Furthermore, the NYM and the scaling approaches have captured the main features of the contact line dynamics. However, we caution that the agreement is not quantitative in all respects. At small $b/R_\infty$, the experimental time constants in Fig. 4(d) deviate from the NYM data and the scaling asymptotics. Several effects may explain this discrepancy: non-linear [28] or spatially non-uniform [29, 30] slip may be operative; the assumption of constant microscopic contact angle [17, 18], $\theta(t) = \theta_\infty$, could be relaxed as well (Fig. S3); furthermore, in ultra-thin ($\sim 5\,\mathrm{nm}$) dewetting polymer films, thermal fluctuations are known to enhance the dynamics [64], which could also affect the effective liquid mobility at the solid-liquid boundary, or at the three-phase contact line. Additionally, at relatively small slip length, other mechanisms of CLM may contribute [16]. The present results thus open a perspective to more detailed studies from both theoretical and experimental sides.

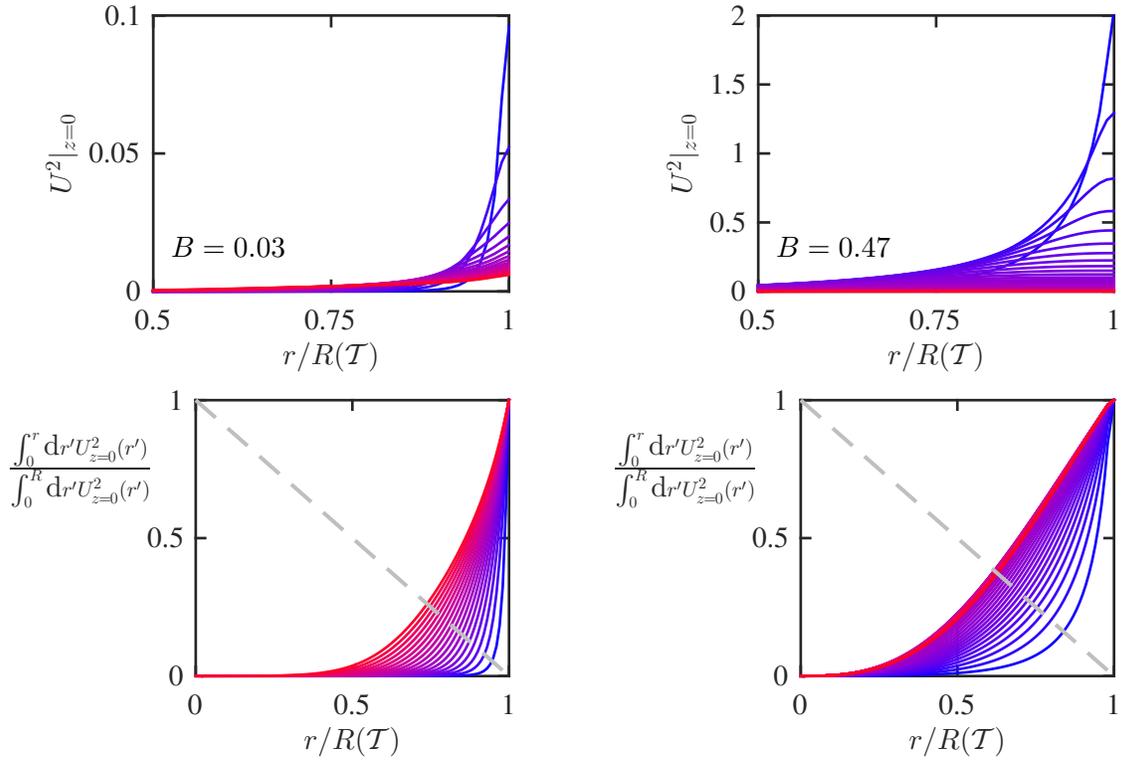

FIG. S5. (top) Squared slip velocity (*i.e.* the fluid velocity at the substrate) as a function of the radial position, normalized by the contact line radius. The dimensionless slip lengths are the same as those for the droplets shown in Fig. 2 of the main text. (bottom) Normalized frictional dissipation power (*i.e.* the dissipation due to fluid sliding against the solid substrate) on a disk of size $r$ as a function of the normalized radial coordinate. The dashed line represents $y = 1 - x$ with $y$ the vertical coordinate and $x$ the horizontal one. The points at which the solid lines cross the dashed one are always below $y = 0.5$, which demonstrates that most of the frictional dissipation occurs near the contact line for both cases, especially at early times (blue).

## ASYMPTOTIC EARLY-TIME SCALING

Here, we provide quantitative arguments for the scaling law given in Eq. (2), using lubrication theory. Since the experiments are performed in a situation where the slip length $b$ is comparable to the initial droplet radius $R_0$, and since we focus on the short-term asymptotic behaviour near the contact line, we invoke the 2D thin-film equation for intermediate slip [65] :

$$\partial_t h + \frac{\gamma b}{\eta} \partial_x \left( h^2 \partial_x^3 h \right) = 0 \; , \tag{S1}$$

where $\gamma$ is the PS-air surface tension and $\eta$ the PS shear viscosity. This PDE describes the visco-capillary evolution of a thin liquid film of profile $h(x,t)$ in space $x$ (coordinate taken in the lab frame) and time $t$, with intermediate slip at the substrate. The contact line is located at $x = R(t)$, and the liquid film is chosen to occupy the $x < R(t)$ region, for instance. The two boundary conditions at the contact line are thus: $h(R(t),t) = 0$, and $\partial_x h(R(t),t) = -\tan\theta_\infty$. We thus assume an equilibrium microscopic contact angle $\theta_\infty$ at any time, consistent with the Navier-Young Model used in the numerics. Let us non-dimensionalize the problem through $x = X R_0$, $t = T \eta R_0^2 / \gamma b$, $R(t) = \mathcal{R}(T) R_0$, $h(x,t) = H(X,T) R_0$, such that

$$\partial_T H + \partial_X \left( H^2 \partial_X^3 H \right) = 0 \; . \tag{S2}$$

We now look for a self-similar asymptotic form at short times near the contact line:

$$H(X,T) = T^\alpha F(\tilde{U}) \; , \tag{S3}$$

$$\tilde{U} = \frac{X - \mathcal{R}(T)}{T^\beta} \; , \tag{S4}$$

where $\alpha$ and $\beta$ are unknown exponents, and $\mathcal{R}(T)$ is the law of interest. Invoking the boundary conditions above, one obtains $\alpha = \beta$, $F(0) = 0$, and $F'(0) = -\tan\theta_\infty$. Then, injecting the self-similar form into Eq. (S2), one gets

$$\alpha(F - \tilde{U}F') = \dot{\mathcal{R}} T^{1-\alpha} F' - T^{1-2\alpha} (F^2 F''')' \; , \tag{S5}$$

where the prime represents the derivative with respect to $\tilde{U}$, and the dot represents the derivative with respect to $T$. The proposed self-similar form is a possible solution if and only if the previous equation is an ODE on the single variable $\tilde{U}$. One possibility is thus given by $\alpha = 1/2$, together with

$$\mathcal{R} = 1 - C\sqrt{T} \; , \tag{S6}$$

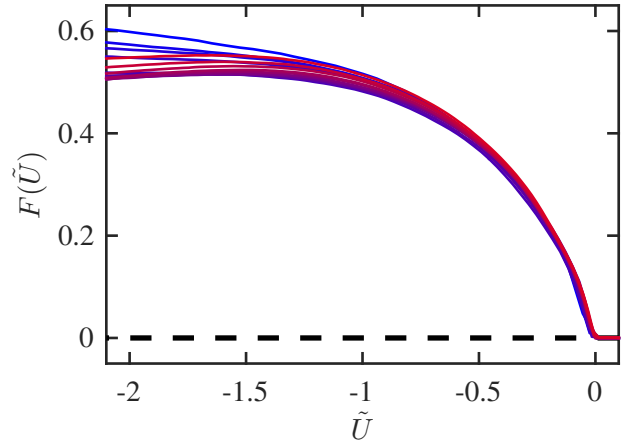

FIG. S6. Self-similar representation (Eqs. (S3) and (S4)) of early-time profiles for a 10 kg/mol PS droplet at 110 °C dewetting on a Si substrate coated with OTS, for times 20 min $< t <$ 160 min.

that satisfies the initial condition, where $C$ is a positive numerical constant, and where the minus sign corresponds to the dewetting situation. In that case, the ODE satisfied by $F$ reads:

$$F + (C - \tilde{U})F' + 2(F^2 F''')' = 0 \ . \tag{S7}$$

The latter could be solved numerically by adding three boundary conditions – especially in the unperturbed far-field region – and by shooting on $C$. Finally, putting back real dimensions, one obtains

$$R_0 - R(t) = C\sqrt{\frac{\gamma b t}{\eta}} \ , \tag{S8}$$

which demonstrates the scaling law of Eq. (2) in the main text. Furthermore, one can validate the proposed self-similar form by comparison with the experimental profiles, as shown in Fig. S6.

We note that the essential ingredients in both the power balance and the thin-film equation are the same: capillarity drives a viscous flow and the dissipation is mostly due to friction at the substrate. The fact that we neglected the $H^3$ term in the thin-film equation, according to the intermediate-slip model [65], is consistent with the fact that we also neglected the bulk viscous dissipation near the contact line in the power balance leading to the scaling of Eq. (2).

## ASYMPTOTIC LATE-TIME SCALINGS

We will assume that the shape of the contracting droplet is close to that of a spherical cap. In the late-time regime of the contraction process, the apparent contact angle $\theta$ (evaluated through the assumed spherical cap profile) deviates from the microscopic contact angle $\theta_\infty$, but the difference $\theta - \theta_\infty$ is small and approaches nil at asymptotically large times.

The interfacial energy of a spherical cap with an apparent contact angle $\theta$ is given by

$$E = \gamma(A_{\mathrm{lv}} - \cos\theta_\infty A_{\mathrm{sl}}) \ , \tag{S9}$$

where we have defined the area of the substrate wet by liquid,

$$A_{\mathrm{sl}} = \pi R^2 \ , \tag{S10}$$

and the interfacial area between the liquid spherical cap and the vapor phase,

$$A_{\mathrm{lv}} = \frac{2\pi(1 - \cos\theta)R^2}{\sin^2\theta} \ . \tag{S11}$$

The contact line radius $R$ is linked to the apparent contact angle $\theta$ by the condition that the volume of the liquid,

$$\Omega = \frac{\pi(1 - \cos\theta)^2(2 + \cos\theta)R^3}{3\sin^3\theta} \ , \tag{S12}$$

is constant.

We first invert Eq. (S12) to obtain

$$R = \left[\frac{3\Omega\sin^3\theta}{\pi(1 - \cos\theta)^2(2 + \sin\theta)}\right]^{1/3} \ , \tag{S13}$$

and insert the latter into Eqs. (S10) and (S11). We thus obtain from Eq. (S9) the total interfacial energy of the spherical cap $E(\theta)$ as a function of the apparent contact angle $\theta$ only. We find

$$\frac{\mathrm{d}E}{\mathrm{d}\theta}\Big|_{\theta=\theta_\infty} = 0 \ , \tag{S14}$$

as expected in mechanical equilibrium for $\theta = \theta_\infty$, and a second derivative

$$\frac{\mathrm{d}^2E}{\mathrm{d}\theta^2}\Big|_{\theta=\theta_\infty} = \frac{2\pi R_\infty{}^2}{2 + \cos\theta_\infty} > 0 \ , \tag{S15}$$

since the spherical cap is a stable shape.

Similarly, we could employ an energy $\tilde{E}(R)$ that depends only on the radius of the contact line $R$ and $\theta(R)$. The results are

$$\frac{\mathrm{d}\tilde{E}}{\mathrm{d}R}\Big|_{R=R_\infty} = 0 \ , \tag{S16}$$

and

$$\frac{\mathrm{d}^2\tilde{E}}{\mathrm{d}R^2}\Big|_{R=R_\infty} = 2\pi\,(2+\cos\theta_\infty)\sin^2\theta_\infty > 0 \ . \tag{S17}$$

Here, we make use of the chain rule of differentiation:

$$\frac{\mathrm{d}\tilde{E}}{\mathrm{d}R}\Big|_{R=R_\infty} \equiv \frac{\mathrm{d}E}{\mathrm{d}\theta}\Big|_{\theta=\theta_\infty}\left(\frac{\mathrm{d}R}{\mathrm{d}\theta}\right)^{-1}\Big|_{\theta=\theta_\infty} \ , \tag{S18}$$

and the identity

$$\frac{\mathrm{d}^2\tilde{E}}{\mathrm{d}R^2}\Big|_{R=R_\infty} \equiv \frac{\mathrm{d}^2E}{\mathrm{d}\theta^2}\Big|_{\theta=\theta_\infty}\left(\frac{\mathrm{d}R}{\mathrm{d}\theta}\right)^{-2}\Big|_{\theta=\theta_\infty} \ , \tag{S19}$$

which holds for an equilibrium state at $\theta=\theta_\infty$ and $R=R_\infty$.

Expanding the energy $\tilde{E}(R)$ around $R=R_\infty$ up to second order yields

$$\tilde{E} = \tilde{E}_\infty + \pi\gamma(2+\cos\theta_\infty)\sin^2\theta_\infty\,(R-R_\infty)^2 \ , \tag{S20}$$

which, when differentiated with respect to $R-R_\infty$, yields a linear restoring force as claimed in the main text. The injected power from the gain of interfacial energy then reads

$$\mathcal{P}_{\mathrm{inj}} = 2\pi\gamma(2+\cos\theta_\infty)\sin^2\theta_\infty\,(R-R_\infty)\,\dot{R} \ . \tag{S21}$$

---

**Weak slip:** $b \ll R$

In the limit of weak slip, the dissipation is mainly due to viscous losses in the bulk flow close to the contact line. In this contact line region, the free surface profile of the droplet is assumed to be a wedge described by: $z = h(x,t) = x\theta$, where $x$ denotes the distance from the contact line. When the droplet spreads with a contact line velocity $\dot{R}$, there is a flow in the wedge that is described by a Poiseuille velocity profile:

$$v = \frac{3\dot{R}}{2h^2}z(z-2h) \ , \tag{S22}$$

with a local average velocity $\left(\int_0^h dz\ v\right)/h = \dot{R}$, and where we assumed a no-slip boundary condition at the substrate and a no-shear boundary condition at the free surface. The partial-slip boundary condition gives a noticeable departure from the Poiseulle profile only in the vicinity of the contact line where $h(x) \lesssim \theta\ell_{\mathrm{m}}$, with a microscopic length, $\ell_{\mathrm{m}} = \tilde{C}\,b$, where $\tilde{C} \sim \mathcal{O}(1)$. Therefore, the viscous dissipation power reads:

$$\mathcal{P}_{\mathrm{dis}} = 2\pi\eta R_\infty\int_{\ell_{\mathrm{m}}}^{R_\infty} dx\int_0^h dz\ (\partial_z v)^2 = \frac{6\pi\eta R_\infty \dot{R}^2}{\theta_\infty}\ln\left(\frac{R_\infty}{\tilde{C}\,b}\right) \ . \tag{S23}$$

Balancing the dissipated power, Eq. (S23), and the injected power Eq. (S21) yields a differential equation for $R(t)$ with the solution

$$R(t) - R_\infty \sim \exp\left(-\frac{t}{\tau}\right) \,,$$
(S24)

with a time constant

$$\tau = \frac{\eta R_\infty}{\gamma} f\left(\frac{b}{R_\infty}, \theta_\infty\right)$$
(S25)

$$f(b/R_\infty) = -\frac{3\ln(b/R_\infty)}{\theta_\infty(2 + \cos\theta_\infty)\sin^2\theta_\infty} + C_1 \,,$$
(S26)

where $C_1$ is a non-universal constant that depends on the global interfacial geometry. Identifying Eq. (S26) with the corresponding expression in the main text, $f = C_0 \ln(b/R_\infty) + C_1$, and substituting $\theta_\infty = 62°$, we find $C_0 = 1.44$.

### Strong slip: $b \gg R$

Because of incompressibility we have $\mathrm{Tr}(\dot{\boldsymbol{\epsilon}}) = 0$ for the rate-of-strain tensor $\dot{\boldsymbol{\epsilon}}$. For flat droplets $\theta_\infty \ll 1$, we can assume an axially symmetric and uniform straining flow in the horizontal and vertical direction of the droplet. In particular, we find that the rate of strain $\dot{\epsilon}_z$ into axial direction is twice the rate-of-strain in the two directions parallel to the substrate:

$$\dot{\epsilon}_x = \dot{\epsilon}_y = -\frac{\dot{\epsilon}_z}{2} \,.$$
(S27)

For a Newtonian fluid, we have $\boldsymbol{\sigma} = \eta \dot{\boldsymbol{\epsilon}}$ which provides us with the viscous dissipation:

$$\mathcal{P}_{\mathrm{dis}} = \int \mathrm{d}V \, \boldsymbol{\sigma} : \dot{\boldsymbol{\epsilon}} \,,$$
(S28)

which yields

$$\mathcal{P}_{\mathrm{dis}} = 12\,\eta\,\dot{\epsilon}^2\,\Omega \,,$$
(S29)

where the volume $\Omega$ is given by Eq. (S12). Balancing the injected power of Eq. (S21) with the dissipated power of Eq. (S29), and relations $\dot{R}/R = \dot{\epsilon}_x = \dot{\epsilon}_y$, we finally arrive at the relaxation timescale:

$$\tau = \frac{\eta R_\infty}{\gamma} \, \frac{2(1 - \cos\theta_\infty)^2}{\sin^5\theta_\infty} \,,$$
(S30)

valid in the limit of flull slip $b/R_\infty \to \infty$. For the particular case considered here, we have

$$\tau(62°) \approx 1.05 \, \frac{\eta R_\infty}{\gamma} \,.$$
(S31)



\end{document}